\documentclass[a4paper,12pt]{article}
\usepackage{amsmath, bm}
\usepackage{pstricks}
\usepackage{pst-node}
\usepackage[ansinew]{inputenc}
\usepackage{amssymb,amsmath}
\usepackage{amsfonts}
\usepackage{epsfig}

\usepackage{pst-node}

 \let\~\widetilde
\def\BSE{\begin{subequations}}\def\ESE{\end{subequations}}
\let \= \equiv

\def\p{\partial}
\def\px{\partial_x}
\def\py{\partial_y}
\def\a{\alpha}
\def\b{\beta}
\def\g{\gamma}
\def\d{\delta}
\def\o{\omega}
\def\wt{\widetilde}
\def\ms{\medskip}
\def\e{\varepsilon}

\font\Sets=msbm10

\def\Integer {\hbox{\Sets Z}}    \def\Real {\hbox{\Sets R}}

\def\Complex {\hbox{\Sets C}}   \def\Natural {\hbox{\Sets N}}

\def\be{\begin{equation}}       \def\ba{\begin{array}}

\def\ee{\end{equation}}         \def\ea{\end{array}}

\def\bea {\begin{eqnarray}}      \def\eea {\end{eqnarray}}

\def\bean{\begin{eqnarray*}}    \def\eean{\end{eqnarray*}}

\def\pa  {\partial}             \def\ti  {\widetilde}

\def\la  {\lambda}

\def\eps{\varepsilon}           \def\ph{\varphi}

\def\const {\mathop{\rm const}\nolimits}

\def\tr    {\mathop{\rm trace}\nolimits}

\def\res   {\mathop{\rm res}  \limits}

\def\diag  {\mathop{\rm diag} \nolimits}

\def\im    {\mathop{\rm Im}   \nolimits}

\def\ker  {\mathop{\rm Ker} \nolimits}

\def\RA {\ \Rightarrow\ }         \def\LRA {\ \Leftrightarrow\ }

\def\qed   {\vrule height0.6em width0.3em depth0pt}

\def\<{\langle} \def\({\left(}  \def\>{\rangle} \def\){\right)}

\def\defeq {\stackrel{\mbox{\rm\small def}}{=}}

\author{Elena Kartashova(1), Alexey Kartashov(2)\\
(1)RISC, J. Kepler University, Linz, Austria\\
(2)AK-Soft, Linz, Austria\\
e-mails: lena@risc.uni-linz.ac.at, alexkart1@gmx.at}

\title{LAMINATED WAVE TURBULENCE: GENERIC ALGORITHMS III}

\begin{document}
%generates titel
\date{}
\maketitle

\begin{abstract}
Model of laminated wave turbulence allows to study statistical and
discrete layers of turbulence in the frame of the same model.
Statistical layer is described by Zakharov-Kolmogorov energy spectra
in the case of irrational enough dispersion function. Discrete layer
is covered by some system(s) of Diophantine equations while their
form is determined by wave dispersion function. This presents a very
special computational challenge - to solve Diophantine equations in
many variables, usually 6 to 8, in high degrees, say 16, in integers
of order $10^{16}$ and more. Generic
 algorithms for solving this problem in the case of
 {\it irrational} dispersion function have been
presented in our previous papers. In this paper we present a new
algorithm for the case of {\it rational} dispersion functions.
Special importance of this case is due to the fact that in wave
systems with rational dispersion the statistical layer does not
exist and the general energy transport is  governed by the discrete
layer alone.


\end{abstract}

%\maketitle

\def\p{\partial}
\def\px{\partial_x}
\def\py{\partial_y}
\def\a{\alpha}
\def\b{\beta}
\def\g{\gamma}
\def\d{\delta}
\def\o{\omega}
\def\wt{\widetilde}
\def\ms{\medskip}

\font\Sets=msbm10

\def\Integer {\hbox{\Sets Z}}    \def\Real {\hbox{\Sets R}}

\def\Complex {\hbox{\Sets C}}   \def\Natural {\hbox{\Sets N}}

\def\be{\begin{equation}}       \def\ba{\begin{array}}

\def\ee{\end{equation}}         \def\ea{\end{array}}

\def\bea {\begin{eqnarray}}      \def\eea {\end{eqnarray}}

\def\bean{\begin{eqnarray*}}    \def\eean{\end{eqnarray*}}

\def\pa  {\partial}             \def\ti  {\widetilde}

\def\la  {\lambda}

\def\eps{\varepsilon}           \def\ph{\varphi}

\def\const {\mathop{\rm const}\nolimits}

\def\tr    {\mathop{\rm trace}\nolimits}

\def\res   {\mathop{\rm res}  \limits}

\def\diag  {\mathop{\rm diag} \nolimits}

\def\im    {\mathop{\rm Im}   \nolimits}

\def\ker  {\mathop{\rm Ker} \nolimits}

\def\RA {\ \Rightarrow\ }         \def\LRA {\ \Leftrightarrow\ }

\def\qed   {\vrule height0.6em width0.3em depth0pt}

\def\<{\langle} \def\({\left(}  \def\>{\rangle} \def\){\right)}

\def\defeq {\stackrel{\mbox{\rm\small def}}{=}}

\newpage

\section{Introduction}
The general theory of fluid mechanics begins in 1741 with the work
of Leonhard Euler who was invited by Frederick the Great to
construct an intrinsic system of water fountains. Euler began with
deducing the equations which are now called Euler equations; they
describe the ideal (inviscid) liquid and are derived from the
classical Newton's conservation laws written for a fluid particle.
Euler equations, regarded with various boundary conditions and
specific values of some parameters describe enormous number of wave
systems, for instance, capillary waves, surface water waves,
atmospheric planetary waves, drift waves in plasma, Tsunami, freak
waves, etc. The general form of reduced Euler equations suitable for
studying one specific type of waves can be written as
$$\mathcal{L}(\varphi)=-\e \mathcal{N}(\varphi)$$
where $\mathcal{L}$ and $\mathcal{N}$ are linear and nonlinear
operators correspondingly, and $\e$ is a small parameter chosen
according to the properties of the wave system under consideration.
For instance, it can be taken as a ratio of wave amplitude to its
length, or as a ratio of a particle velocity to the phase velocity,
or some other way. A linear wave is then a solution of the
corresponding linear equation $\mathcal{L}(\psi)=0$ and has standard
form $ A \exp {i[ \vec{k}\vec{x} - \o t]} $ with amplitude $A$, wave
vector $ \vec{k}$ and dispersion function $\o_i=\o(\vec{k}_i)$. The
form of dispersion function is defined then by boundary conditions.
The existence of a small parameter $\e$ allows to reduce the study
of  all nonlinear waves to those which are resonantly interacting,
that is, satisfy resonant conditions
\begin{eqnarray}\label{open}
\begin{cases}
\omega (\vec k_1) \pm \omega (\vec k_2)\pm ... \pm \omega (\vec k_{n+1}) = 0,\\
\vec k_1 \pm \vec k_2 \pm ... \pm \vec k_{n+1} = 0.
\end{cases}
\end{eqnarray}
Notice that amplitudes of resonantly interacting waves are not
constant any more and standard multi-scale method yields the
corresponding system of ordinary differential equations (ODEs) on
these amplitudes. The energetic behavior of a wave system depends
drastically on whether wave vectors $\vec k_i$ have real or integer
coordinates. The first case (real-valued coordinates) is treated in
the frame of statistical wave turbulence (SWT) theory \cite{lvov},
with additional assumption that $\omega (\vec k_i)/\omega (\vec
k_j)$ is an algebraic number of degree $\ge 2$. The energy transport
in these systems is covered by the wave kinetic equation. The second
case (integer-valued coordinates) is described by discrete wave
turbulence (DWT) theory \cite{KarAll}, and energy transport is
presented by a few quasi-periodic processes. Model of laminated
turbulence\cite{lam} presents SWT and DWT as two layers of a wave
system, with elaborate transition from one layer to another. One of
the novel problems emerging from this model is the necessity to
solve
(\ref{open}) for very big integers.\\

In the first two articles\cite{kk2006-1},\cite{kk2006-2} of this
series we presented algorithms for finding resonant wave
interactions for {\it irrational} dispersion functions, with two
illustrative examples: (1) gravitational water waves,
$\o=\sqrt[4]{m^2+n^2}$ (4-wave interactions); and (2) ocean
planetary waves, $\o=1/\sqrt{m^2+n^2}$ (3-wave interactions). The
key points of the presentation were, first, that our algorithms for
these cases differ only in some details and their core is applicable
to a wide class of dispersion functions, thus justifying the name of
"generic". Second, irrational equations in integers were solved
without use of floating-point arithmetic and not even resolving the
irrationalities involved. This gave us an enormous gain both in
performance time and orders of numbers used.\\

 In the present paper we construct a special algorithm
 for solving (\ref{open}) in case of a
{\it rational} dispersion function. Notice that for any rational
dispersion function, $\omega (\vec k_i)/\omega (\vec k_j)$ is
obviously a rational number, that is, an algebraic number of degree
1. It makes SWT theory not applicative for these type of wave
systems because statistical layer of turbulence {\it does not
exists} and the whole energetic behavior is covered by the discrete
layer only. This makes the creation of some fast algorithm for
computing integer solutions of (\ref{open}) for the case of rational
dispersion function of high importance.

\section{General idea of the algorithm}
Obviously,  any equation in rational functions in integers can be
trivially transformed into a Diophantine equation. For%%
\be \label{ratfun} \sum_i{\frac{P_i}{Q_i}}=0 \ee%%
the corresponding Diophantine equation will be \be \label{diafun}
\sum_i{(P_i\prod_jQ_j)}=0 \quad j\ne i\ee%%
which, however, leads to huge powers and extensive search. The idea
underlying our algorithm is quite simple and we illustrate it by the
example below.

\paragraph{Example} Suppose we need solve in integers an equation
\be \label{ratfunex}a=b\frac{P}{Q}, \quad 0<a \le a_0, 0<b \le b_0
\ee%%
where P/Q is an irreducible fraction. We could transform it into
$aQ=bP$ and perform exhaustive search in the region $0<a<a_0,
0<b<b_0$ with computational complexity $O(a_0b_0)$.

However, we notice that the number $b\frac{P}{Q}$ is integer only if
$b$ is a multiple of the denominator $Q$. Then $(a,b)$ is a solution
only if $b=kQ$ with integer $k$. Which immediately gives $a=kP$ and
$(kP, kQ)$ is a solution for any $k, \quad 1 \le k \le \min(P/a_0,
Q/b_0)$ and these are all the solutions of the equation. Notice that
there is no search {at all} involved.

 To show the power of the approach outlined above in practice, we
 proceed further with the example of spherical planetary waves.

\subsection{Example 1: spherical planetary waves}
The turbulence of the spherical planetary waves is governed by the
barotropic vorticity equation on a sphere
\begin{equation}\label{BVE}
\frac{\partial \triangle \psi}{\partial t} + 2 \frac{\partial
\psi}{\partial \lambda} +  J(\psi,\triangle  \psi) =0
\end{equation}
where
$$
\triangle \psi = \frac{\partial^2 \psi}{\partial \phi^2}+ \frac
{1}{\cos^2 \phi} \frac {\partial^2 \psi}{\partial \lambda^2} - \tan
\phi \frac{\partial \psi}{\partial \phi} \quad \mbox{and} \quad
J(a,b)= \frac{1}{cos \phi}(\frac{\partial a}{\partial \lambda}
\frac{\partial b}{\partial \phi} - \frac{\partial a}{\partial \phi}
\frac{\partial b}{\partial \lambda}).$$ A linear wave has
 the form
\begin{equation} \label{math_shock}
A P_n^m (\sin \phi) \exp{i[m \lambda +\frac{2m}{n(n+1)}
t]},\nonumber
\end{equation}
with constant wave amplitude $A$,  dispersion function $\ \omega =
-2m /[n(n+1)]\ $ and $ \ P_n^m (x) \ $ being the associated Legendre
function of degree $n$ and order $m$. Resonance conditions in this
case have form\cite{ped}:%%
 \bea\label{prosetdef1} \begin{cases}
\o_1 + \o_2 = \o_3 \\
m_1+m_2=m_3 \\
m_i \le n_i \quad \forall i=1,2,3 \\
|n_1-n_2| \le n_3 \le n_1+n_2 \\
n_1+n_2+n_3 = 1(\mod 2) \\
n_i \ne n_j \quad \forall i \ne j \\
\end{cases}\eea%%
where $\o_i=m_i/{(n_i(n_{i+1}))}$ .\\

We are going to find all the solutions of Sys. (\ref{prosetdef1}) in
a finite domain $D$, i.e. $0 < m_i, n_i \le D \quad \forall
i=1,2,3$. In our numerical experiments we operated with $D=1000$
further called the main domain.

\subsection{Computational Preliminaries}

The straightforward approach would be to multiply the first equation
of Sys. (\ref{prosetdef1}) with all three denominators
$n_i(n_{i+1})$, substitute $m_3$ with $m_1+m_2$ and perform full
search on $m_1, m_2, n_1, n_2, n_3$. This evidently implies $D^5$
computation time and operating with numbers of the order of $D^5$.
For the main domain $D=1000$ this is halfway feasible with a large
computer but clearly not for everyday use with a usual PC. Moreover,
computation time {\it and} order of numbers used grow rapidly with
the domain, so when need for computations in larger domains arises,
as it surely will, the algorithm will fail.\\

We are going to present a far more efficient algorithm.

\subsection{Algorithm Description}
\begin{itemize}
\item{} Step 1: Search on $n_1, n_2, n_3.$\\

 The search on $n_1, n_2, n_3$ is
organized conventionally. Without loss of generality, consider $n_1
< n_2$. Notice that $n_3$ always lies between $n_1$ and $n_2$ and
>from the two "triangle inequalities" of Sys. (\ref{prosetdef1}) the
second one always holds, while the first one implies $n_3 > n_2-n_1$
which limits the search on $n_3$ if $n_2-n_1>n_1$. The oddity
condition allows us to run the cycle on $n_3$ in steps of 2.

Up to now, the computational complexity is $O(D^3)$.\\

\item{} Step 2: Cycles elimination on $m_1, m_2$.\\

The numbers of the form $n(n+1)$ are sometimes called "box numbers"
(analogous to the square numbers $n^2$) and we introduce notation
$b_i=n_i(n_i+1)$. Now we rewrite the first equation of Sys.
(\ref{prosetdef1}) as%%
\be \label{prodef1} m_1/b_1+m_2/b_2=m_1/b_3+m_2/b_3 \ee%%
or%%
 \be \label{profr} m_2 = m_1 \cdot \frac{b_3-b_1}{b_2-b_3}\cdot
\frac{b_2}{b_1} \ee%%
 Let us find the greatest common divisor GCD of
the numerator and denominator of the fraction on  the right side and
reduce by it. The equation now has the form \be \label{profrred} m_2
= m_1 \cdot \frac{R_N}{R_D} \ee and every solution has the form \be
\label{prosol} m_1=kR_D, m_2 = kR_N,  \quad k \le min(n_1/R_D,
n_3/(R_N+R_D)).\ee The second condition follows from \be
\label{profrred} m_3 = m_1 + m_2 \le n_3 \ee and is stronger than
$m_2 \le n_2$.

The computational complexity of the whole algorithm is thus $O(\log
D D^3)$, $D^3$ for the cycle on $n_i$ and $\log D$ for the GCD.

\paragraph{Remark} The algorithm above implies operating with
numbers of the order of $D^4$ in one certain place, namely,
transforming \be \label{remmain} \frac{b_3-b_1}{b_2-b_3}\cdot
\frac{b_2}{b_1} \Rightarrow \frac{R_N}{R_D}. \ee This could lead to
overflows be $D$ large and computer small, say $D=1000$ and 32 bit
computer or $D=10^6$ and 64 bit computer. There
is, however, an elegant way to avoid difficulties at this point
which we describe in the next Step.\\

\item{} Step 3: Avoiding multiplications.\\

 Given the fraction product, we first reduce $b_3-b_1$ and
$b_2-b_3$ by their GCD, then $b_2$ and $b_1$ by their GCD. This
leaves us with a product of two irreducible fractions
$(r_{31}/r_{23})\cdot(r_2/r_1)$. Now we reduce crosswise: $r_{31}$
and $r_1$, $r_{23}$ and $r_2$. The last reduction gives an
"irreducible product" of two fractions
$(rr_{31}/rr_{23})\cdot(rr_2/rr_1)$, i.e. had we performed the
multiplications, the resulting fraction would stay irreducible. The
reduction schema is presented in Fig. \ref{f:red}.

\begin{figure}[h]
\centerline{\psfig{file=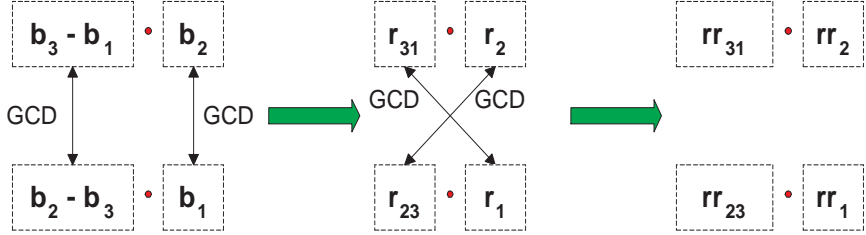, width=12cm, height=4cm}}
\vspace*{8pt} \caption{Bringing a product of two fractions to
complete irreducibility without multiplying. \label{f:red}}
\end{figure}

We still do not perform multiplications for fear of an overflow. But
now it is evident that a solution can only exist if $rr_{23} \le
n_1, \quad rr_1 \le n_1, \quad rr_{31} \le n_2, \quad rr_2 \le n_2$.
We first check these inequalities; if one or more of them do not
hold, we proceed with the $n$-cycle, otherwise we may safely perform
multiplications (both products do not exceed $D^2$) and look for
solutions.
\end{itemize}

\subsection{Example 2: drift waves in a channel}
The turbulence of the drift waves is described by the same equation
as in Sec.2.1 but in Descartes  coordinates and in the infinite
channel\cite{bnz}. In this case dispersion function has a (slightly
simplified) form $\o=2m/(n^2+1)$ and resonance conditions are%%
 \bea\label{prosetdef2} \begin{cases}
\o_1 + \o_2 = \o_3 \\
m_1+m_2=m_3 \\
m_i \le n_i \quad \forall i=1,2,3 \\
n_i \ne n_j \quad \forall i \ne j \\
\end{cases}\eea%%

Search cycles on $n_1$, $n_2$, $n_3$ become somewhat more extensive
due to the lack of the two conditions mentioned above. On the other
hand, the core of the algorithm - the four reductions (Step 2) - are
preserved one-to-one, as well as the post-reduction overflow check
(Step 3).

The computational complexity of the whole algorithm is also $O(\log
D D^3)$ as in the previous case.

\section{Numerical results and some discussion}
Our algorithm has been implemented in VBA programming language; for
$D=1000$ computation time (without disk output of solutions found)
on a low-end PC (800 MHz Pentium III, 512 MB RAM) is about 7.5
minutes. Altogether 7282 solutions (example 1) have been found.
Some overall numerical data is given in the Tables and Figures below.\\

\begin{figure}[h]
\centerline{\psfig{file=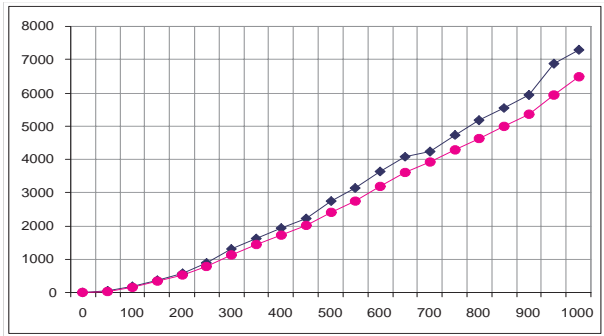, width=8cm, height=4.5cm}}
\vspace*{8pt} \caption{Example 1: Number of all solutions in partial
domains: squares (points with diamonds) and circles (points with
circles).\label{f:ex1_sol}}
\end{figure}

\begin{figure}[h]
\centerline{\psfig{file=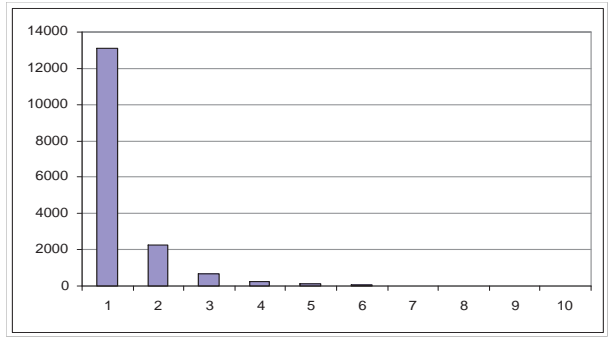, width=8cm, height=4.5cm}}
\vspace*{8pt} \caption{Example 1: Histogram of vector
multiplicities. \label{f:ex1_hist}}
\end{figure}

In Fig.\ref{f:ex1_sol} the number of solutions in partial domains is
shown for the first example (atmospheric planetary waves) and we
conclude that the solutions are concentrated along $X$ and $Y$ axes.
In Fig.\ref{f:ex1_hist} the histogram of vector multiplicities is
presented which shows in how many solutions one vector can
participate. On the axis $X$ the multiplicity of a vector is shown
and on the axis $Y$ the number of vectors with a given multiplicity.
One can see immediately that most part of vectors take part only in
one solution and multiplicity
decreases exponentially with the number of solutions.\\

As to our second example (drift  waves in a channel) we notice,
first of all, much less solutions ($477$) in the same main domain
$D=1000$. Therefore, not much can be said about the asymptotic of
solution number in partial domains. There is no need to present
multiplicities graphically in this case. In the whole calculation
domain $D=1000$ there is just one vector $(1,5)$ participating in
solutions with multiplicity $5$, one vector $(78, 99)$ with
multiplicity $4$ and the overall
distribution is as follows:\\

\begin{tabular}{|c|c|c|c|c|c|}
\hline
    Multiplicity & 1 & 2 & 3 & 4 & 5 \\
\hline
    Number of vectors & 1254 & 72 & 8 & 1 & 1\\
\hline
\end{tabular}\\

\noindent {\small Table 1. Example 2: Vector multiplicities.}\\

\begin{figure}[h]
\centerline{\psfig{file=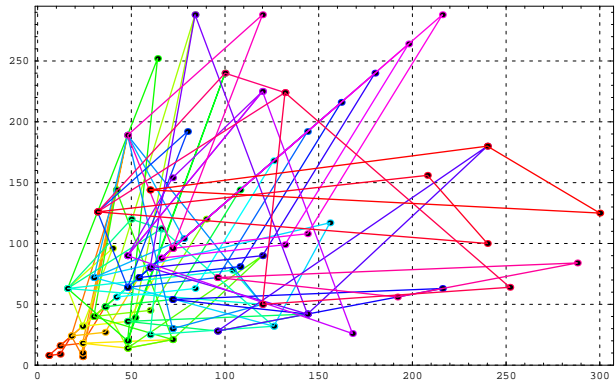, width=8cm,
height=5cm}} \vspace*{8pt} \caption{Example of geometrical structure
of a solution set. \label{f:geom}}
\end{figure}

In order to understand the energetic behavior of 2-dimensional
discrete wave system, the standard way is to present is graphically
on the integer lattice in following way. Each node with coordinates
$m,n$ presents a corresponding wave vector $\vec k=(m,n)$ and
nodes-vectors are connected by lines is they are parts of the same
solution. An example of this {\it geometrical} structure is given in
Fig.\ref{f:geom}.\\

This geometrical representation is needed in order to understand
what sort of equations (ODEs) for the amplitudes of resonantly
interacting waves we have to solve. Namely, one single triangle in
$\vec k$-space corresponds to%%
 \bea \label{sing}
 \begin{cases}\dot{A}_1= \a_1 A_2A_3  \\
 \dot{A}_2= \a_2 A_1A_3  \\
 \dot{A}_3= \a_3 A_1A_2
 \end{cases}
\eea
 where coefficients $\a_i$ are known functions on $m_i, n_i, \
i=1,2,3$. If one wave takes part in two solutions, we get two
systems of this form connected {\it via} this wave, for instance, if
the second solution corresponds to $$
 \dot{A}_4= \a_4 A_5A_6 , \quad
 \dot{A}_5= \a_5 A_4A_6 , \quad
 \dot{A}_6= \a_6 A_4A_4 ,
$$ and they are connected {\it via} one wave, say $A_3=A_4$, then
corresponding system of ODEs takes form %%
\bea \label{but}
\begin{cases}
 \dot{A}_1= \a_1 A_2A_3  \\
 \dot{A}_2= \a_2 A_1A_3 \\
 \dot{A}_3= \frac{1}{2}(\a_3A_1A_2 +\a_4A_5A_6)\\
 \dot{A}_5= \a_5 A_3A_6  \\
 \dot{A}_6= \a_6 A_3A_5
\end{cases}
\eea
and so on.\\

\begin{figure}[h]
\centerline{\psfig{file=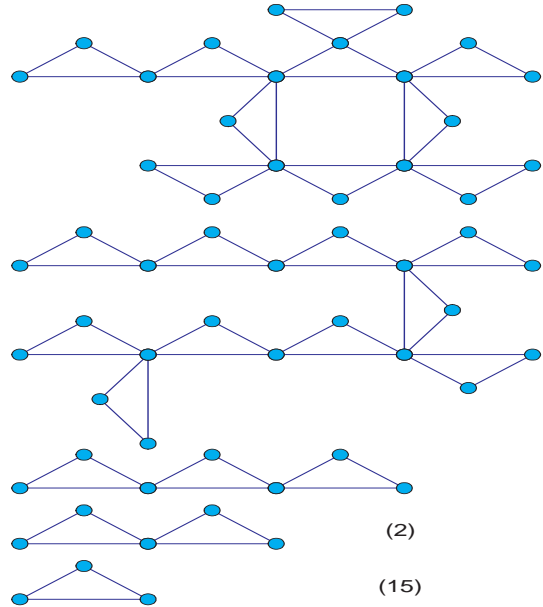, width=7cm, height=8cm}}
\vspace*{8pt} \caption{Example 1: Topological structure of solutions
for $D=50$. \label{f:ex1_top}}
\end{figure}

Obviously, the geometrical structure is too confusing to be
informative and what we really need is a {\it topological} structure
of a solution set, i.e. the graph formed by triangles as {\it
primary elements}. Namely, it is enough to compute all {\it
non-isomorphic} topological elements because all isomorphic elements
are described by the same system of ODEs. For instance, all primary
elements (isolated resonant triads) are described by (\ref{sing}),
all "butterflies" (groups of two connected triads) are described by
(\ref{but}), etc. The difference between two isomorphic topological
elements lies in the coefficients $\a_i$ which are functions of the
wave numbers, and therefore, will take different magnitudes for
different resonant triads. Topological structure of the solution set
for our first example is shown in Fig. \ref{f:ex1_top} for domain
$m,n \le 50.$ This domain contains 42 solutions: 15 isolated
triangles, two "butterflies" (groups of two connected triangles),
one  chain of three connected triangles and
two more complex graphs.\\

\section{Summary}
This paper concludes the series of three papers on generic
algorithms for laminated wave turbulence. We have presented
algorithms for polynomial dispersion function depending irrationally
on the wave vector length $k=\sqrt{m^2+n^2}$ and for an arbitrary
rational dispersion function. We have also shown that the
topological elements of the solution set (for the discrete layer of
turbulence) give the whole information about the energy transport in
these wave systems. In fact we applied this approach for our first
example (atmospheric planetary waves) and studied all the
topological elements in the meteorologically  significant domain
($m,n \le 21$) for climate range processes\cite{KL1}. More
precisely, we have found analytically solutions of corresponding
systems of the form (\ref{sing}) in terms of Jacobean elliptic
functions and computed their periods and other properties for
characteristic meteorological data. As a result, a novel model of
the known physical phenomena - intra-seasonal oscillations in the
Earth atmosphere - has been
developed.\\

Our further interest lies now in the area of symbolic computations.
Indeed, beginning with an equation like (\ref{BVE}) with given
boundary conditions, we have a completely constructive procedure of
obtaining precise form for: I) resonance conditions (\ref{open});
II) coefficients $\a_i$ of the primary element (\ref{sing}); III)
all topological elements - as graphs and as corresponding systems of
ODEs. All this can be programmed symbolically in MATHEMATICA and
solutions can be found using our generic algorithms. This is an
on-going work now and we plan to create a useful program tool for
making basic research in the area of discrete wave turbulence. There
are some open mathematical questions yet to be solved - for
instance, the problem of graph isomorphism appearing at the step
when all different topological elements have to be computed.\\

Another possible development would be the study of the 4-wave
interactions, that is, with primary elements being not triads but
quartets of waves (see example of gravitational water
waves\cite{kk2006-1},\cite{kk2006-2}). The same constructive
procedure as for 3-wave interactions can be applied but the
resulting topology will be much more complicated due to the
principal difference between 3- and 4-wave systems. In 3-wave system
there exist the only mechanism for the energy transport - transport
over the scales. In 4-wave system there are two qualitatively
different mechanisms of the energy flow - over the scales and over
the phases, and they can combine in a highly nontrivial
way.\\

 {\bf Acknowledgement}.  E.K.  acknowledges the support of the Austrian Science
Foundation (FWF) under projects SFB F013/F1304.

\end{document}